\documentclass[aps,prb,twocolumn]{revtex4-1}

\usepackage[pdftex]{graphics}
\usepackage{amssymb,amsmath}
\usepackage[usenames,dvipsnames]{color}

\begin{document}
\newcommand{\red}[1]{{\color{red}#1}}

\title{Magnetoelectroluminescence in organic light emitting diodes}
\author{Joseph E. Lawrence}
\affiliation{Department of Chemistry, University of Oxford, Physical and Theoretical Chemistry Laboratory, South Parks Road, Oxford, OX1 3QZ, UK}
\author{Alan M. Lewis}
\affiliation{Department of Chemistry, University of Oxford, Physical and Theoretical Chemistry Laboratory, South Parks Road, Oxford, OX1 3QZ, UK}
\author{David E. Manolopoulos}
\affiliation{Department of Chemistry, University of Oxford, Physical and Theoretical Chemistry Laboratory, South Parks Road, Oxford, OX1 3QZ, UK}
\author{P. J. Hore}
\affiliation{Department of Chemistry, University of Oxford, Physical and Theoretical Chemistry Laboratory, South Parks Road, Oxford, OX1 3QZ, UK}

\begin{abstract}
The magnetoelectroluminescence of conjugated organic polymer films is widely accepted to arise from a polaron pair mechanism, but their magnetoconductance is less well understood. Here we derive a new relationship between the experimentally measurable magnetoelectroluminescence and magnetoconductance and the theoretically calculable singlet yield of the polaron pair recombination reaction. This relationship is expected to be valid regardless of the mechanism of the magnetoconductance, provided the mobilities of the free polarons are independent of the applied magnetic field {(i.e., provided one discounts the possibility of spin-dependent transport)}. We also discuss the semiclassical calculation of the singlet yield of the polaron pair recombination reaction for materials such as poly(2,5-dioctyloxy-paraphenylene vinylene) (DOO-PPV), the hyperfine fields in the polarons of which can be extracted from light-induced electron spin resonance measurements. The resulting theory is shown to give good agreement with experimental data for both normal (H-) and deuterated (D-) DOO-PPV over a wide range of magnetic field strengths once singlet-triplet dephasing is taken into account. Without this effect, which has not been included in any previous simulation of magnetoelectroluminescence, it is not possible to reproduce the experimental data for both isotopologues in a consistent fashion. Our results also indicate that the magnetoconductance of DOO-PPV cannot be solely due to the effect of the magnetic field on the dissociation of polaron pairs. 
\end{abstract}

\maketitle

\section{Introduction}

Electroluminescence is an important and much studied property of semiconducting films of conjugated organic polymers which underlies their commercial application in organic light emitting diodes (OLEDs).\cite{Yang97,Hoofman98,Grozema02} There is therefore significant interest in understanding the mechanism of this electroluminesence and the factors that affect it. 

The observation of magnetoelectroluminescence (MEL) -- that is, a change in the electroluminescence upon application of a magnetic field -- has provided direct evidence for the polaron pair mechanism described in Section~II.\cite{Lupton10,Ehrenfreund12} However, the physical interactions which govern the spin dynamics of the polaron pair are less well understood. Recent isotopic substitution experiments have strongly suggested that hyperfine interactions between the electron spin and the spins of the hydrogen nuclei in the polaron pair play a crucial role,\cite{Nguyen10a,Nguyen10b} but it remains unclear whether other physical effects are also important.

Previous analyses of MEL have all been undertaken with an implicit assumption that the number density of polaron pairs in the material is independent of the applied magnetic field.\cite{Ehrenfreund12,Nguyen10a,Kersten11} This leads to a very simple relationship between the experimentally accessible magnetoelectroluminescence and the theoretically accessible singlet yield of the polaron pair recombination reaction (see Section~III). However, it is not an assumption which is easily justified. The fact that magnetoconductance (MC) is also well known in OLEDs implies that the steady state number density of free polarons, and therefore also of polaron pairs, must be a function of the applied magnetic field, and it seems likely that this is the reason why previous analyses have been unable to account for the form of the experimental MEL at high magnetic field strengths\cite{Ehrenfreund12,Nguyen10a,Kersten11} 

In Section~IV, we present a new and more general relationship between the MEL and the singlet yield of the polaron pair recombination reaction which avoids the assumption of constant polaron pair number density. This relationship involves the measured MC, but it does not make any assumptions about its underlying mechanism, {which may have a variety of different contributions in addition to the polaron pair mechanism itself.\cite{Frankevich92,Hu07,Bobbert07,Lupton08,Janssen11,Cox14}} 

Section~V goes on to discuss the calculation of the singlet yield of the polaron pair recombination reaction using the early semiclassical theory of Schulten and Wolynes,\cite{Schulten78} which is expected to be reasonably reliable for polarons with $N\sim 100$ nuclear spins. The only input that this theory requires beyond the overall singlet and triplet decay rates of the polaron pair is the standard deviation $B_{\rm hyp}$ of the hyperfine field in each polaron. Section~VI argues that this is the only experimentally accessible piece of information about the hyperfine interactions in conjugated organic polarons, because their electron spin resonance (ESR) spectra do not exhibit resolved hyperfine splittings.\cite{Kuroda00,Zezin04} All that is actually seen in these spectra is a broad first-derivative line which integrates to give a Gaussian profile with a full width at half maximum (FWHM) proportional to $B_{\rm hyp}$.

Section~VII returns to the question of whether hyperfine and Zeeman interactions are enough to account for the spin dynamics in polaron pairs, using the experimental MEL and MC data of Nguyen {\em et al.}\cite{Nguyen10a,Nguyen10b} for both normal (H-) and deuterated (D-) poly(2,5-dioctyloxy-paraphenylene vinylene) (DOO-PPV) for illustration. We argue on the basis of the Weller equation,\cite{Weller83} and the anticipated effect of deuteration on the polaron hyperfine fields, that these data cannot be explained in terms of hyperfine and Zeeman interactions alone:  there must also be some other physical effect influencing the spin dynamics. 

\begin{figure}[t]
\centering
\resizebox{0.8\columnwidth}{!} {\includegraphics{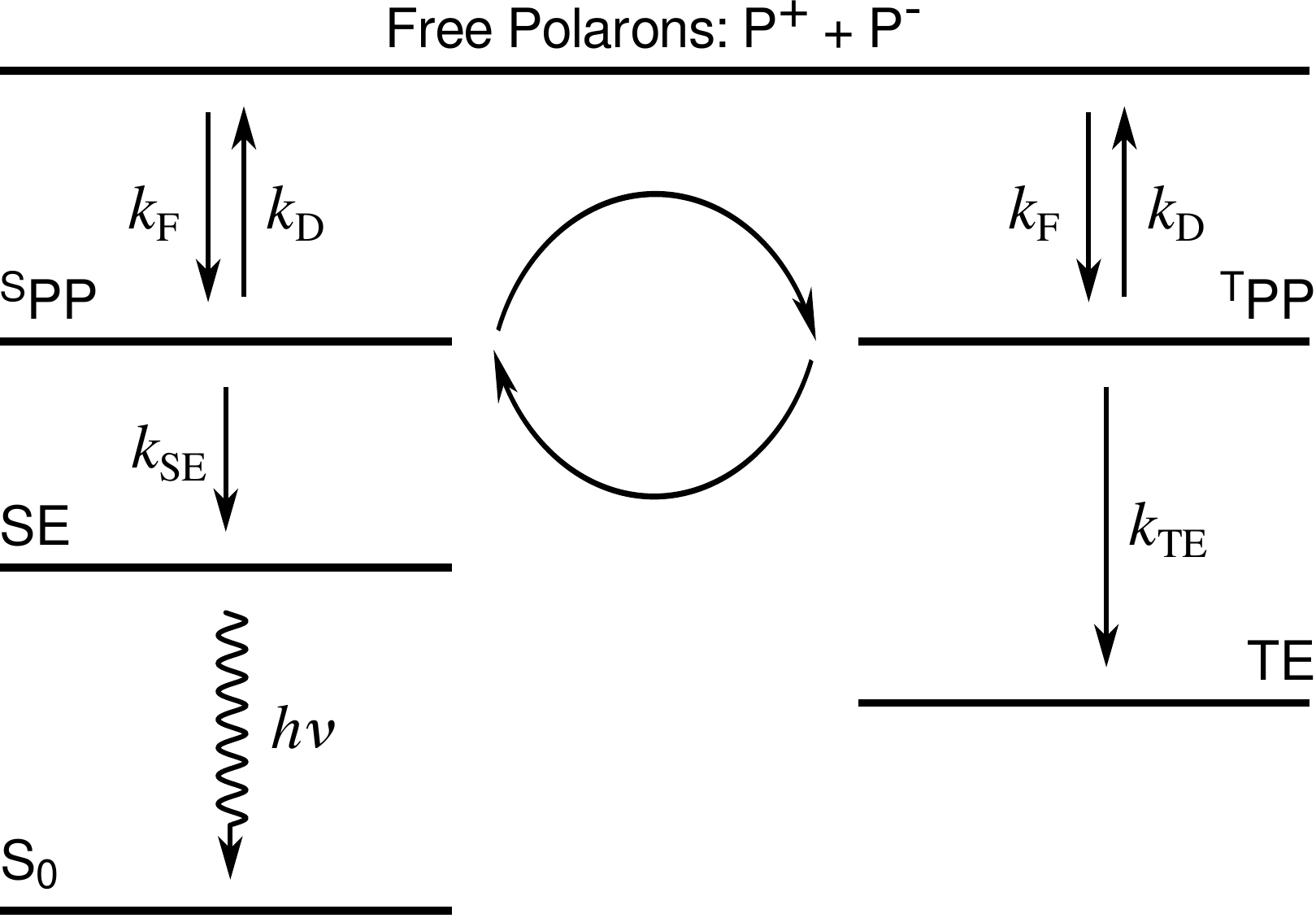}}
\caption{Mechanism of electroluminescence. P$^+$ and P$^-$ are free positively and negatively charged polarons. $^{\rm S}$PP and $^{\rm T}$PP are Coulombically bound pairs of polarons on adjacent polymers in the singlet and triplet states, respectively. These can interconvert by hyperfine and Zeeman-mediated intersystem crossing, represented by the semicircular arrows. SE and TE refer to singlet and triplet excitons of a single polymer, formed by electron transfer from P$^-$ to P$^+$ in the polaron pair. The former of these can fluoresce (luminesce) to the ground S$_0$ state. The singlet and triplet formation rate constants of the polaron pair are assumed to be the same ($k_{\rm F}$), as are their rate constants for dissociation into free polarons ($k_{\rm D}$), although this latter assumption could be removed without affecting any of the present results. The total singlet and triplet decay rate constants of the polaron pair are $k_{\rm S} = k_{\rm SE}+k_{\rm D}$ and $k_{\rm T} = k_{\rm TE}+k_{\rm D}$.}
\label{PPdiagram}
\end{figure}

Several possible physical effects have already been considered and discounted for polaron pairs in conjugated organic polymers like DOO-PPV.\cite{Ehrenfreund12} For example, the spin-orbit interaction is unlikely to play a significant role in mediating the intersystem crossing in a polymer without any heavy atoms, and whatever small difference there may be between the g-values of the two electrons in the polaron pair is unlikely to be significant at the magnetic field strengths of Nguyen {\em et al.}'s experiments. However, comparatively little attention has yet been given in this context to the role of electron spin relaxation. We suggest in Section~VII that the experimental data for both isotopologues of DOO-PPV might be explained by including the effect of \lq\lq singlet-triplet dephasing\rq\rq,\cite{Pope99,Breuer06} which could arise (for example) from the modulation of the exchange interaction between the electrons in the polaron pair as the polarons migrate along and between their polymer chains.\cite{Willard15} 

This suggestion is backed up by explicit simulations of H-DOO-PPV and D-DOO-PPV in Section~VIII, which show that the experimental data of Nguyen {\em et al.}\cite{Nguyen10a,Nguyen10b} can be reproduced almost quantitatively for both isotopologues over a wide range of magnetic field strengths when singlet-triplet dephasing is taken into account. Section~IX concludes the paper with a discussion of the kinetic parameters that are obtained from our analysis of DOO-PPV and some comments about the implications of our results for the mechanism of its magnetoconductance.

\section{The polaron pair mechanism}

It is widely believed that electroluminescence in organic polymer devices occurs through the polaron pair mechanism illustrated in Fig.~1.\cite{Ehrenfreund12,Lupton10} The injection of electrons and holes into the polymer film leads to the formation of positive and negative polarons (P$^+$ and P$^-$). These diffuse through the material under the influence of the applied potential until they encounter a polaron of opposite charge on an adjacent polymer chain, with which they form a loosely bound polaron pair (PP). Since the spins of the polarons are initially uncorrelated, singlet and triplet polaron pairs are formed in a statistical ratio of 1:3. A singlet (or triplet) polaron pair may then form an intra-chain exciton with a rate constant $k_{\rm SE}$ ($k_{\rm TE}$), with the electron hopping onto the polymer chain on which the hole resides or vice versa. Alternatively, the polaron pair could dissociate to reform free polarons with a rate constant $k_{\rm D}$, which we have assumed to be the same for both the singlet and triplet states. If the total decay rate constants $k_{\rm S} = k_{\rm SE}+k_{\rm D}$ and $k_{\rm T} = k_{\rm TE}+k_{\rm D}$ are small enough, significant interconversion between the singlet and triplet states of the polaron pair is possible, due to the hyperfine and Zeeman interactions within each polaron. If $k_{\rm S} \neq k_{\rm T}$, then changes in the frequency of interconversion will affect the singlet exciton yield. Since only the singlet exciton is emissive, and the strength of an applied magnetic field $B$ alters the rate of interconversion between the singlet and triplet states, the polaron pair mechanism gives rise to magnetoelectroluminescence,
\begin{equation}
{\rm MEL}(B) = {{\rm EL}(B)\over {\rm EL}(0)}-1, 
\end{equation}
where ${\rm EL}(B)$ is the electroluminescence in the presence of the field and ${\rm EL}(0)$ is the electroluminescence in its absence.

\begin{figure}[t]
\centering
\resizebox{0.4\columnwidth}{!} {\includegraphics{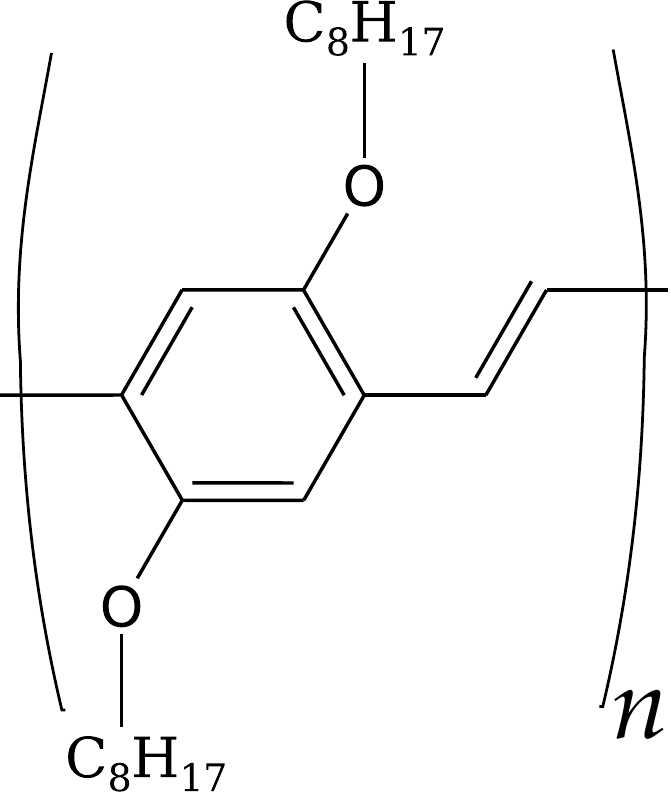}}
\caption{The repeat unit of DOO-PPV [poly(2,5-dioctyloxy-paraphenylene vinylene)].}
\label{DOO-PPV}
\end{figure}

A magnetic field effect on the conductance has also been observed in organic polymer devices. However, the origin of this effect is far less well understood, and it is a subject of continuing debate.\cite{Frankevich92,Bobbert07,Cox14,Lupton08,Hu07,Janssen11} We shall sidestep this issue here by making no assumptions about the mechanism of the magnetoconductance,
\begin{equation}
{\rm MC}(B) = {{\rm C}(B)\over {\rm  C}(0)}-1, 
\end{equation}
where ${\rm C}(B)$ is the conductance in the presence of the field and ${\rm C}(0)$ is the conductance in its absence. 

Both the MEL and MC have been measured in a variety of materials. In what follows, we will concentrate on DOO-PPV, whose repeat unit is shown in Fig.~2. This material was chosen because the impact of deuteration on its magnetic field effects has been studied experimentally,\cite{Nguyen10a,Nguyen10b} providing direct evidence for the hyperfine-mediated intersystem crossing between the singlet and triplet states of its polaron pairs. 

\begin{figure}[t]
\centering
\resizebox{0.9\columnwidth}{!} {\includegraphics{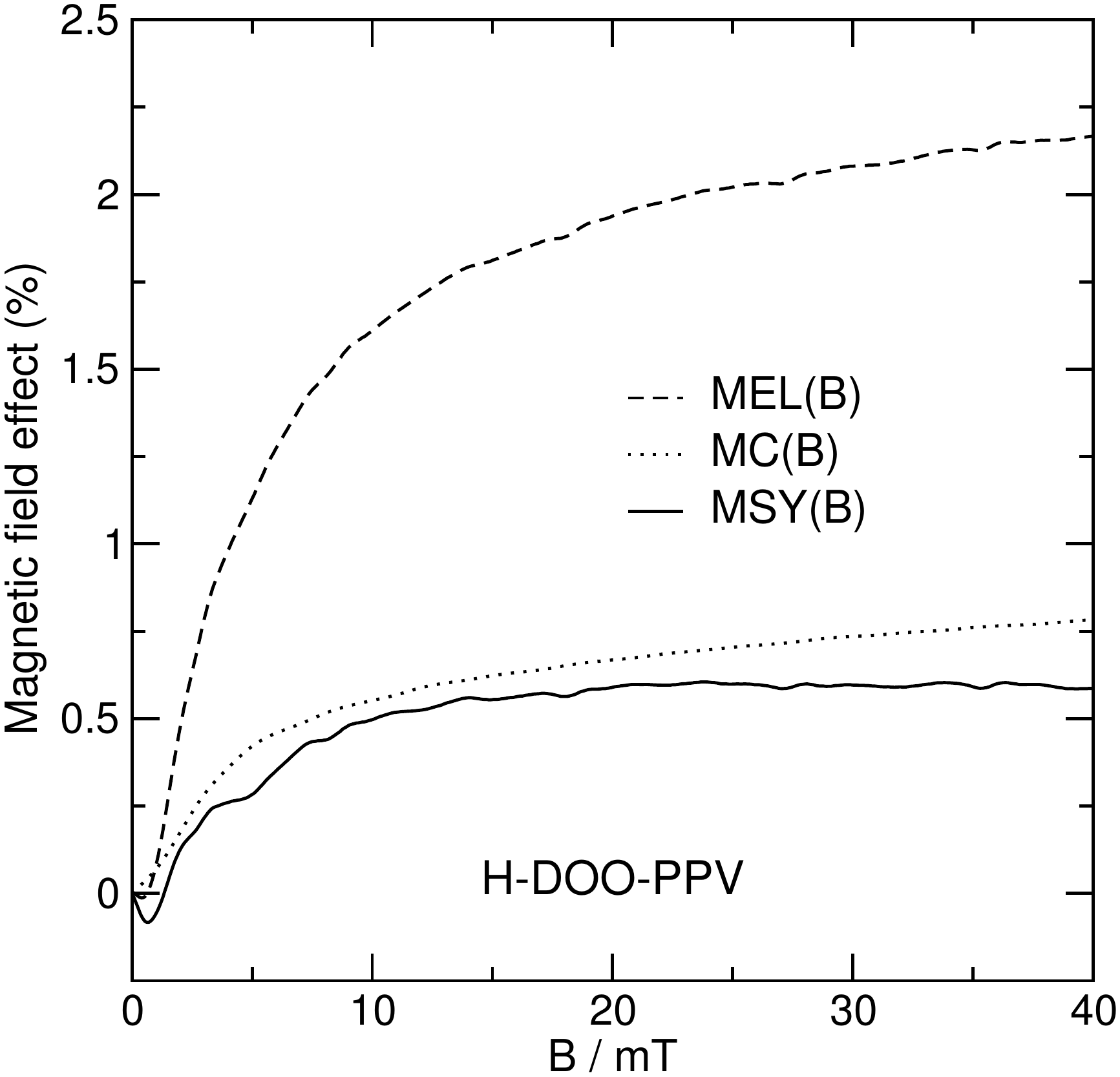}}
\caption{Comparison of the experimental magnetic field effects on the electroluminescence (MEL) and conductance (MC) of H-DOO-PPV, taken from Refs.~\onlinecite{Nguyen10a} and~\onlinecite{Nguyen10b}. Also shown is the magnetic field effect on the singlet yield of the polaron pair recombination reaction (MSY), as given by the relationship in Eq.~(24).}
\label{Experimental}
\end{figure}

The experimentally determined MEL and MC in undeuterated DOO-PPV (H-DOO-PPV) are shown in Fig.~3, and those in deuterated DOO-PPV (D-DOO-PPV) in Fig.~4. In order to compare these experimental observations with theoretical calculations, we shall need to derive a new relationship between the observed MEL and MC and the singlet yield of the polaron pair recombination reaction. This is because, as we have already mentioned in the Introduction and shall discuss in more detail next, we believe there is something wrong with the expression that has been used to calculate MEL($B$) in the past.\cite{Ehrenfreund12,Nguyen10a,Kersten11} 

\begin{figure}[t]
\centering
\resizebox{0.9\columnwidth}{!} {\includegraphics{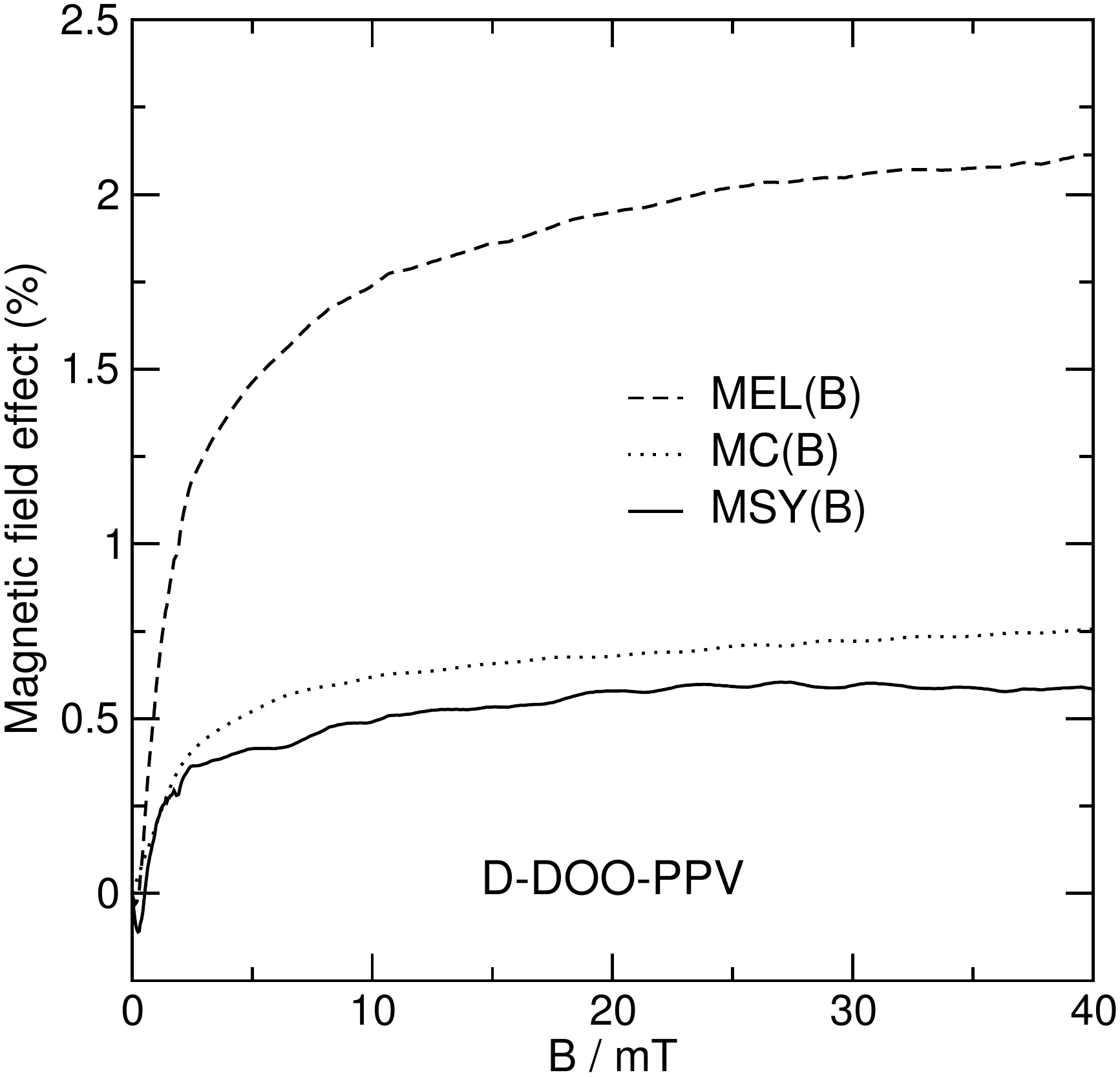}}
\caption{As in Fig.~3, but for D-DOO-PPV. The experimental MEL and MC curves are again from Refs.~\onlinecite{Nguyen10a} and~\onlinecite{Nguyen10b}.}
\label{Experimental}
\end{figure}

\section{The standard approximation to MEL}

Clearly, the electroluminescence at a given magnetic field strength will depend not only on the singlet exciton yield of the polaron pair decay, but also on the steady state concentration of polaron pairs in the polymer film. In all previous work,\cite{Ehrenfreund12,Nguyen10a,Kersten11} an implicit assumption has been made that the steady state concentration of polaron pairs is independent of the applied magnetic field. In that case, one arrives at a straightforward expression for the MEL within the polaron pair model,\cite{Ehrenfreund12,Nguyen10a,Kersten11}
\begin{equation}
{\rm MEL}(B) \simeq \frac{\Phi_{\rm SE}(B)}{\Phi_{\rm SE}(0)}-1,
\end{equation}
where $\Phi_{\rm SE}(B)$ is the singlet exciton yield of the polaron pair recombination reaction when the applied magnetic field has magnitude $B$. Alternatively, since 
\begin{equation}
\Phi_{\rm SE}(B) = {k_{\rm SE}\over k_{\rm S}}\Phi_{\rm S}(B),
\end{equation}
where $\Phi_{\rm S}(B)$ is the total singlet yield of the reaction (the fraction of polaron pairs which, having formed in a statistical 1:3 mixture of singlet and triplet states, either recombine to give singlet excitons or dissociate to give free polaron pairs in the singlet state), one has
\begin{equation}
{\rm MEL}(B) \simeq {\rm MSY}(B) = \frac{\Phi_{\rm S}(B)}{\Phi_{\rm S}(0)}-1.
\end{equation}
That is, under the assumption that the steady state concentration of polaron pairs is independent of the applied magnetic field, the magnetoelectroluminescence is identical to the magnetic field effect on the overall singlet yield of the polaron pair reaction, MSY$(B)$.

However, whenever Eq.~(5) has been used in conjunction with quantum mechanical or semiclassical calculations of $\Phi_{\rm S}(B)$ in the past,\cite{Ehrenfreund12,Nguyen10a,Kersten11} the results have failed to capture the correct qualitative behaviour of the observed MEL$(B)$, which unlike MSY$(B)$ does not plateau at high field strengths (see the experimental results in Figs.~3 and~4 above and the theoretical results in Figs.~6 and~7 below). So let us now develop an alternative expression that allows for the magnetic field dependence of the steady state concentration (or number density) of polaron pairs, $n_{\rm PP}(B)$.

\section{A new relationship between MEL and MC}
 
It not legitimate to apply a simple (incoherent) steady state analysis to the polaron pair mechanism illustrated in Fig.~1, because the hyperfine and Zeeman mediated intersystem crossing between the singlet and triplet states of the polaron pair is a coherent process.  One is therefore obliged to base the analysis on the quantum mechanical density operator of the polaron pair.

Let the normalised density operator of this polaron pair be $\hat{\rho}(B)$, and define the number density operator
$\hat{n}(B) = n_{\rm PP}(B)\hat{\rho}(B)$, so that 
\begin{equation}
{\rm tr}[\hat{n}(B)] = n_{\rm PP}(B){\rm tr}[\hat{\rho}(B)] = n_{\rm PP}(B). 
\end{equation}
Then in view of Fig.~1, the equation that governs the evolution of $\hat{n}(B)$ is
\begin{equation}
{d\over dt}\hat{n}(B) = -\hat{\hat{L}}(B)\hat{n}(B)+k_{\rm F}\,n_+(B)n_-(B)\hat{\rho}_0,
\end{equation}
with $\hat{\rho}_0 = \hat{1}/N$.
Here $\hat{\hat{L}}(B)$ is the Liouville super-operator that accounts for the spin evolution, recombination, dissociation, and spin relaxation processes of the polaron pair, $n_{\pm}(B)$ are the number densities of positively and negatively charged free polarons in the polymer film, $\hat{1}$ is the unit operator on the Hilbert space of the polaron pair, and $N={\rm tr}[\hat{1}]$ is the total number of electron and nuclear spin states in this space (so tr$[\hat{\rho}_0]=1$).

In the absence of electron spin relaxation, the first term on the right hand side of Eq.~(7) can be written out more explicitly as
\begin{equation}
-\hat{\hat{L}}(B)\hat{n}(B) = -i\left[\hat{H}(B),\hat{n}(B)\right]-\left\{\hat{K},\hat{n}(B)\right\}, 
\end{equation}
where $[\hat{A},\hat{B}]$ is a commutator and $\{\hat{A},\hat{B}\}$ is an anti-commutator, $\hat{H}(B)$ is the spin Hamiltonian of the polaron pair (which depends on $B$ because of the electronic Zeeman interaction -- see Eq.~(26) below), and\cite{Haberkorn76}
\begin{equation}
\hat{K}={k_{\rm S}\over 2}\hat{P}_{\rm S}+{k_{\rm T}\over 2}\hat{P}_{\rm T},
\end{equation}
with
\begin{equation}
\hat{P}_{\rm S} = {1\over 4}\hat{1}-\hat{\bf S}_{1}\cdot\hat{\bf S}_2,
\end{equation}
\begin{equation}
\hat{P}_{\rm T} = {3\over 4}\hat{1}+\hat{\bf S}_{1}\cdot\hat{\bf S}_2.
\end{equation}
Here $k_{\rm S}$ and $k_{\rm T}$ are the total first-order rate constants for loss of population from the singlet and triplet states of the polaron pair, $\hat{\bf S}_1$ and $\hat{\bf S}_2$ are the spin angular momentum operators of its two unpaired electrons, and $\hat{P}_{\rm S}$ and $\hat{P}_{\rm T}$ are the projection operators onto the singlet and triplet electronic subspaces of the polaron pair, respectively. 

When the device has reached a steady state (constant current and constant electroluminescence in the presence of the applied magnetic field), $d\hat{n}(B)/dt=0$. So from Eq.~(7), the steady-state number density operator of the polaron pairs is
\begin{equation}
\hat{n}(B) = k_{\rm F}\,n_+(B)n_-(B)\,\hat{\hat{L}}(B)^{-1}\hat{\rho}_0.
\end{equation}
From this, one can calculate the steady-state number density of singlet polaron pairs, $n_{\rm S}(B)$, using the standard formula
\begin{equation}
n_{\rm S}(B) = {\rm tr}[\hat{P}_{\rm S}\,\hat{n}(B)], 
\end{equation}
which gives 
\begin{equation}
n_{\rm S}(B) = {k_{\rm F}\over k_{\rm S}}\,n_+(B)n_-(B)\Phi_{\rm S}(B),
\end{equation}
with
\begin{equation}
\Phi_{\rm S}(B) = k_{\rm S}{\rm tr}[\hat{P}_{\rm S}\hat{\hat{L}}(B)^{-1}\hat{\rho}_0].
\end{equation}

The reason why we have written Eq.~(14) in this way is that $\Phi_{\rm S}(B)$ in Eq.~(15) is precisely the singlet yield of the recombination reaction that would be obtained if the polaron pair were prepared with the initial density operator $\hat{\rho}_0=\hat{1}/N$ at time $t=0$ 
(corresponding to a statistical 1:3 initial population of the singlet and triplet states), 
as can be seen from the standard expression for this singlet yield\cite{Jones10}
\begin{equation}
\Phi_{\rm S}(B) = k_{\rm S}\int_0^{\infty} {\rm tr}\left[\hat{P}_{\rm S}e^{-\hat{\hat{L}}(B)t}\hat{\rho}_0\right]\,dt.
\end{equation}
This therefore establishes a connection between the present steady-state argument and the formula for MEL($B$) in Eq.~(5). 

However, the right-hand side of Eq.~(14) also contains a magnetic field dependence in the factor $n_+(B)n_-(B)$, which suggests that Eq.~(5) is incorrect. Indeed it is clear from Fig.~1 that the steady state fluoresence intensity EL$(B)$ will be proportional to
\begin{equation}
k_{\rm SE}n_{\rm S}(B) = {k_{\rm SE}k_{\rm F}\over k_{\rm S}}\,n_{+}(B)n_{-}(B)\Phi_{\rm S}(B),
\end{equation}
and therefore, from Eq.~(1), that Eq.~(5) should be replaced by
\begin{equation}
{\rm MEL}(B) = {n_+(B)n_-(B)\Phi_{\rm S}(B)\over n_+(0)n_-(0)\Phi_{\rm S}(0)}-1.
\end{equation}

In order to make any further progress, we shall now need to make a simplifying assumption about the relationship between $n_+(B)$ and $n_-(B)$, the steady-state number densities of positively and negatively charged polarons in the polymer film. The assumption we shall make is that the film has approximately Ohmic contacts, in which case there will only be a relatively small net space charge $\delta n(B)=n_+(B)-n_-(B)$ within it.\cite{Parmenter59} I.e., we shall assume that
\begin{equation}
n_{\pm}(B) \simeq n(B),
\end{equation}
where $n(B)=[n_+(B)+n_-(B)]/2$. This approximation is certainly reasonable\cite{Leger03} for dialkoxy PPV-based OLEDs with calcium cathodes and poly(3,4-ethylenedioxythiophene) - poly(styrenesulfonate) (PEDOT-PSS) hole transport layers of the type employed in the  experiments of Nguyen {\em et al.}\cite{Nguyen10a,Nguyen10b} 

With this approximation, Eq.~(18) becomes\cite{MELapproximation}
\begin{equation}
{\rm MEL}(B) \simeq {n(B)^2\Phi_{\rm S}(B)\over n(0)^2\Phi_{\rm S}(0)}-1.
\end{equation}
Assuming further that the mobilities $\mu_+$ and $\mu_-$ of the free polarons (P$^+$ and P$^-$) are independent of the strength of the applied magnetic field {(i.e., neglecting the possibility of spin-dependent transport)}, the magnetoconductance
\begin{equation}
{\rm MC}(B) = {{\mu_{+}n_+(B)+\mu_-n_-(B)}\over{\mu_+n_+(0)+\mu_-n_-(0)}}-1,
\end{equation}
becomes\cite{MCapproximation}
\begin{equation}
{\rm MC}(B) \simeq {n(B)\over n(0)}-1.
\end{equation}
So combining Eqs.~(20) and (22) and rearranging gives
\begin{equation}
[{\rm MEL}(B)+1] \simeq [{\rm MC}(B)+1]^2[{\rm MSY}(B)+1],
\end{equation}
where ${\rm MSY}(B)$ is defined by the equality in Eq.~(5). 

This is our central result -- a new relationship between ${\rm MEL}(B)$ and ${\rm MC}(B)$ that involves the magnetic field effect on the overall singlet yield of the polaron pair recombination reaction, ${\rm MSY}(B)$. The only two assumptions we have made in deriving this result are that the contacts are approximately Ohmic and the mobilities $\mu_+$ and $\mu_-$ of P$^+$ and P$^-$ are independent of the applied magnetic field. Given these assumptions, the result is a direct consequence of the polaron pair mechanism for magnetoelectroluminescence illustrated in Fig.~1. 

Notice in particular that we have not made any assumptions about the mechanism of the magnetoconductance, other than that the applied magnetic field does not change the mobilities of the positively and negatively charged free polarons. This is consistent with many of the mechanisms that have been proposed for magnetoconductance in polymer films, including the polaron pair mechanism,\cite{Frankevich92} the bipolaron mechanism in the regime of positive MC,\cite{Bobbert07} and a more recently proposed mechanism involving the trapping of polarons at defect sites.\cite{Cox14} In all three of these scenarios, the magnetic field effect on the conductance is believed to arise from a change in the free polaron number density in the presence of the magnetic field rather than a change in the free polaron mobility. Indeed it is explicitly stated in Ref.~\onlinecite{Frankevich92} that no magnetic field effect is found experimentally on the mobilities of the free polarons in PPV derivatives.

The curves labelled MSY($B$) in Figs.~3 and~4 were obtained from the experimental MEL($B$) and MC($B$) curves by rearranging Eq.~(23) into the form
\begin{equation}
{\rm MSY}(B) \simeq {[{\rm MEL}(B)+1]\over [{\rm MC}(B)+1]^2}-1.
\end{equation}
Another way of viewing Eq.~(23) is thus that it provides a link between the experimentally measurable magnetoelectroluminescence and magnetoconductance and the theoretically calculable magnetic field effect on the singlet yield of the polaron pair recombination reaction. We shall now take this view and move on to discuss the calculation of $\Phi_{\rm S}(B)$, and hence MSY($B$), for materials like DOO-PPV.

\section{Semiclassical calculation of the singlet yield}

The Hamiltonian $\hat{H}(B)$ in Eq.~(8) that governs the coherent electron spin evolution between the singlet and triplet states of the polaron pair is given to within a good approximation by\cite{Schulten78}
\begin{equation}
\hat{H}(B) = \hat{H}_1(B)+\hat{H}_2(B),  
\end{equation}
where
\begin{equation}
\hat{H}_i(B)=-\gamma_i{B}\hat{S}_{iz} + \sum_{k=1}^{N_i} a_{ik} \hat{\bf I}_{ik}\cdot\hat{\bf S}_i, 
\end{equation}
for $i=1$ (P$^+$) and 2 (P$^-$). Here $\hat{\bf S}_i$ is the electron spin angular momentum operator of the unpaired electron in polaron $i$, $a_{ik}$ is the hyperfine coupling constant between this electron spin and the $k$-th nuclear spin in the polaron, and $\hat{\bf I}_{ik}$ is the corresponding nuclear spin angular momentum operator. In the electronic Zeeman term $-\gamma_iB\hat{S}_{iz}$, $\gamma_i$ is the gyromagnetic ratio of the electron spin, and the magnetic field has been chosen to be in the laboratory $z$ direction.

Note that the Hamiltonian in Eq.~(25) neglects the dipolar and exchange coupling between the two electrons in the polaron pair. These will be comparatively weak in a material such as DOO-PPV, in which the long DOO side chains will tend to keep the polarons on neighbouring polymers quite far apart. It also neglects the Zeeman interactions of the nuclear spins with the applied magnetic field, which will be weaker still because of the small gyromagnetic ratios of the nuclear spins.

In principle, we could now simply insert the Hamiltonian in Eq.~(25) into Eq.~(8), and do the spin dynamics calculation quantum mechanically. However, this would require a detailed knowledge of the hyperfine coupling constants $a_{ik}$ in the two polarons. These hyperfine coupling constants, or rather their distribution, could in principle be calculated from a consideration of the vibronic-coupling and disorder-induced localisation of electrons in positively and negatively charged DOO-PPV polymers,\cite{Marcus14} but such a calculation is well beyond the scope of the present study. And in any case, an exact quantum mechanical spin dynamical calculation of $\Phi_{\rm S}(B)$ would be quite impractical for a polaron with $N_i\sim 100$ nuclear spins.  

To avoid both of these problems, we shall resort here to a simple semiclassical approximation to the spin dynamics proposed by Schulten and Wolynes,\cite{Schulten78} which is expected to be reasonably reliable for a polaron with $N_i\sim 100$ nuclear spins. This approximation neglects the (comparatively slow) nuclear spin evolution in such a polaron by replacing the hyperfine-weighted sum of its nuclear spin operators
\begin{equation}
\hat{\bf h}_i= \sum_{k=1}^{N_i} a_{ik}\hat{\bf I}_{ik} 
\end{equation}
with a Gaussian distribution of static hyperfine fields ${\bf h}_i$,
\begin{equation}
P_i({\bf h}_i) = \left({3\over 2\pi B_{{\rm hyp},i}^2}\right)^{3/2} e^{-3|{\bf h}_i|^2/2B_{{\rm hyp},i}^2}
\end{equation}
where 
\begin{equation}
B_{{\rm hyp},i} = \left<|\hat{\bf h}_i|^2\right>^{1/2} = \sqrt{\sum_{k=1}^{N_i} a_{ik}^2 I_{ik}(I_{ik}+1)}. 
\end{equation}
For each pair of polaron hyperfine fields ${\bf h}_1$ and ${\bf h}_2$, the Hamiltonian in Eqs.~(25) and~(26) is thus replaced with
\begin{equation}
\hat{H}(B,{\bf h}_1,{\bf h}_2) = \boldsymbol{\omega}_1\cdot\hat{\bf S}_1+\boldsymbol{\omega}_2\cdot\hat{\bf S}_2,
\end{equation}
where
\begin{equation}
\boldsymbol{\omega}_i = {\bf h}_i-\gamma_i B \boldsymbol{k} 
\end{equation}
and $\boldsymbol{k}$ is a unit vector in the $z$ direction. When the Liouvillian $\hat{\hat{L}}(B)$ in Eq.~(8) is modified accordingly, to become a function of ${\bf h}_1$ and ${\bf h}_2$ that acts on density operators in the Liouville space of the two electron spins, 
\begin{equation}
-\hat{\hat{L}}(B,{\bf h}_1,{\bf h}_2)\hat{\rho} = -i\left[\hat{H}(B,{\bf h}_1,{\bf h}_2),\hat{\rho}\right]-\left\{\hat{K},\hat{\rho}\right\}, 
\end{equation}
one finds that the Schulten-Wolynes approximation to the singlet yield $\Phi_{\rm S}(B)$ in Eq.~(15) becomes
\begin{eqnarray}
\Phi_{\rm S}(B) \simeq \int P_1({\bf h}_1)\,d{\bf h}_1 \int P_2({\bf h}_2)\,d{\bf h}_2\nonumber\\
\times k_{\rm S}\,{\rm tr}[\hat{P}_{\rm S}\hat{\hat{L}}(B,{\bf h}_1,{\bf h}_2)^{-1}\hat{\rho}_0],
\end{eqnarray}
where $\hat{\rho}_0$ is now $\hat{1}/4$ and the trace is simply over the four states in the Hilbert space of the two electrons. This both makes the calculation of $\Phi_{\rm S}(B)$ straightforward and reduces the hyperfine coupling constants $a_{ik}$ in each polaron to a single physically relevant parameter $B_{{\rm hyp},i}$.

\section{Hyperfine fields in DOO-PPV}

In fact, this single physical parameter is all that can be deduced about the distribution of the hyperfine coupling constants from experimental measurements, because the electron spin resonance (ESR) spectra of condensed phase PPV derivatives do not exhibit any resolved hyperfine splittings.\cite{Kuroda00,Zezin04} All that is seen in these spectra is a broad first-derivative ESR line which integrates to give an approximately Gaussian profile with a full width at half maximum (FWHM) of 
\begin{equation}
{\rm FWHM} = \sqrt{{8 \ln 2\over 3}} B_{{\rm hyp},i},
\end{equation}
as one would expect from the Gaussian distribution of $h_{iz}$ in Eq.~(28). The measured ESR spectra are thus entirely consistent with the Schulten-Wolynes approximation that we have used to obtain Eq.~(33).

In particular, Kuroda {\em et al.}\cite{Kuroda00} have measured the light-induced electron spin resonance spectra of thin films of two different dialkoxy derivatives of PPV (MEH-PPV and CN-PPV). They found a FWHM of $0.66$ mT for one and $0.45$ mT for the other, with no evidence for any difference  between the contributions of positive and negative polarons to either ESR signal. More recently, Zezin {\em et al.}\cite{Zezin04} have measured an ESR linewidth of $0.5$ mT for the positive polarons in long oligomers of DOO-PPV in an irradiated glassy toluene solution at 77 K in the presence of an electron scavenger, and performed similar experiments in the presence and absence of the scavenger to deduce that the positive and negative polarons of MEH-PPV have somewhat different ESR linewidths (0.37 mT and 0.59 mT, respectively).

Since all of these linewidths are fairly similar, and since the results we shall present below are fairly insensitive to the precise choice of $B_{{\rm hyp},i}$, we shall avoid introducing too many free parameters into our calculations by assuming a FWHM ESR linewidth of 0.5 mT for both the positive and the negative polarons in thin films of H-DOO-PPV. The use of 0.5 mT for the positive polarons is consistent with the long DOO-PPV oligomer experiments of Zezin {\em et al.}\cite{Zezin04} The assumption that this is the same for both positive and negative polarons is consistent with the thin film experiments of Kuroda {\em et al.},\cite{Kuroda00} and with the particle-hole symmetry of the Pariser-Parr-Pople Hamiltonian for PPV that these authors used to interpret their results.\cite{Shimoi95,ParticleHole}

According to Eq.~(34), a FWHM of 0.5 mT gives $B_{\rm hyp}({\rm H})=0.37$ mT for H-DOO-PPV, and the corresponding parameter for D-DOO-PPV can be worked out as follows. Aside from $^{13}$C nuclei, which will be present throughout the polymer with $\sim 1$\% natural abundance, the only magnetic nuclei in H-DOO-PPV are protons.  Since H has $I=1/2$ and D has $I=1$, and the hyperfine coupling constants $a_{ik}$ in Eq.~(29) are proportional to the gyromagnetic ratios $\gamma_{\rm H}$ and $\gamma_{\rm D}$ of the two nuclei, the effect of deuteration will be to reduce $B_{\rm hyp}$ to{\cite{Ratioof4}} 
\begin{equation}
B_{\rm hyp}({\rm D}) = {\gamma_{\rm D}\over\gamma_{\rm H}}\sqrt{8\over 3}\,B_{\rm hyp}({\rm H})\simeq {1\over 4}B_{\rm hyp}({\rm H}),
\end{equation}
which gives $B_{\rm hyp}({\rm D}) = 0.093$ mT. 

To summarise, the magnetic field effect on the singlet yield of the polaron pair recombination reaction, MSY$(B)$, can be calculated semiclassically from Eqs.~(5) and~(33). The second of these equations involves the hyperfine field strengths $B_{{\rm hyp},i}$ of the positive and negative polarons P$^+$ and P$^-$ ($i=1$ and 2) in the pair. These hyperfine fields have yet to be measured directly in an ESR experiment on thin films of DOO-PPV itself, so we have suggested plausible values on the basis of related experiments that have been performed on thin films of similar dialkoxy PPVs and on solutions of oligomers of DOO-PPV. 

\section{Singlet-triplet dephasing}

Let us now explain why hyperfine interactions alone are not enough to account for the experimental data in Figs.~3 and~4. The MSY($B$)s for H-DOO-PPV and D-DOO-PPV in these figures are compared on an expanded scale in Fig.~5. Both curves tend to the same asymptote MSY($\infty$) at high field strengths, but they differ in the intermediate field region where the magnetic field effect is still growing. It is convenient to characterise each curve in this region with a single number $B_{1/2}$, the magnetic field strength at which MSY($B$) has reached half its asymptotic value. If the observed $B_{1/2}$ values were solely due to the hyperfine fields in the two polarons, $B_{{\rm hyp},1}$ and $B_{{\rm hyp},2}$, one would expect them to conform to the Weller equation\cite{Weller83}
\begin{equation}
B_{1/2} = 2{{B_{{\rm hyp},1}^2+B_{{\rm hyp},2}^2}\over {B_{{\rm hyp},1}+B_{{\rm hyp},2}}},
\end{equation}
which reduces to $B_{1/2}=2B_{\rm hyp}$ when the hyperfine fields of the positive and negative polarons are the same. However, the $B_{1/2}$ values in Fig.~5 clearly do not satisfy this equation. From the experimental data, $B_{1/2}({\rm H})\simeq 5.3$ mT and $B_{1/2}({\rm D})\simeq 2.0$ mT, both of which are more than an order of magnitude larger than our ESR-based estimates of the hyperfine field strengths $B_{\rm hyp}({\rm H})\simeq 0.37$ mT and $B_{\rm hyp}({\rm D})\simeq 0.093$ mT. Moreover the ratio $B_{1/2}({\rm H})/B_{1/2}({\rm D})\simeq 2.65$ is inconsistent with the ratio $B_{\rm hyp}({\rm H})/B_{\rm hyp}({\rm D)}\simeq 4$ one would predict from Eq.~(35).

\begin{figure}[t]
\centering
\resizebox{0.9\columnwidth}{!} {\includegraphics{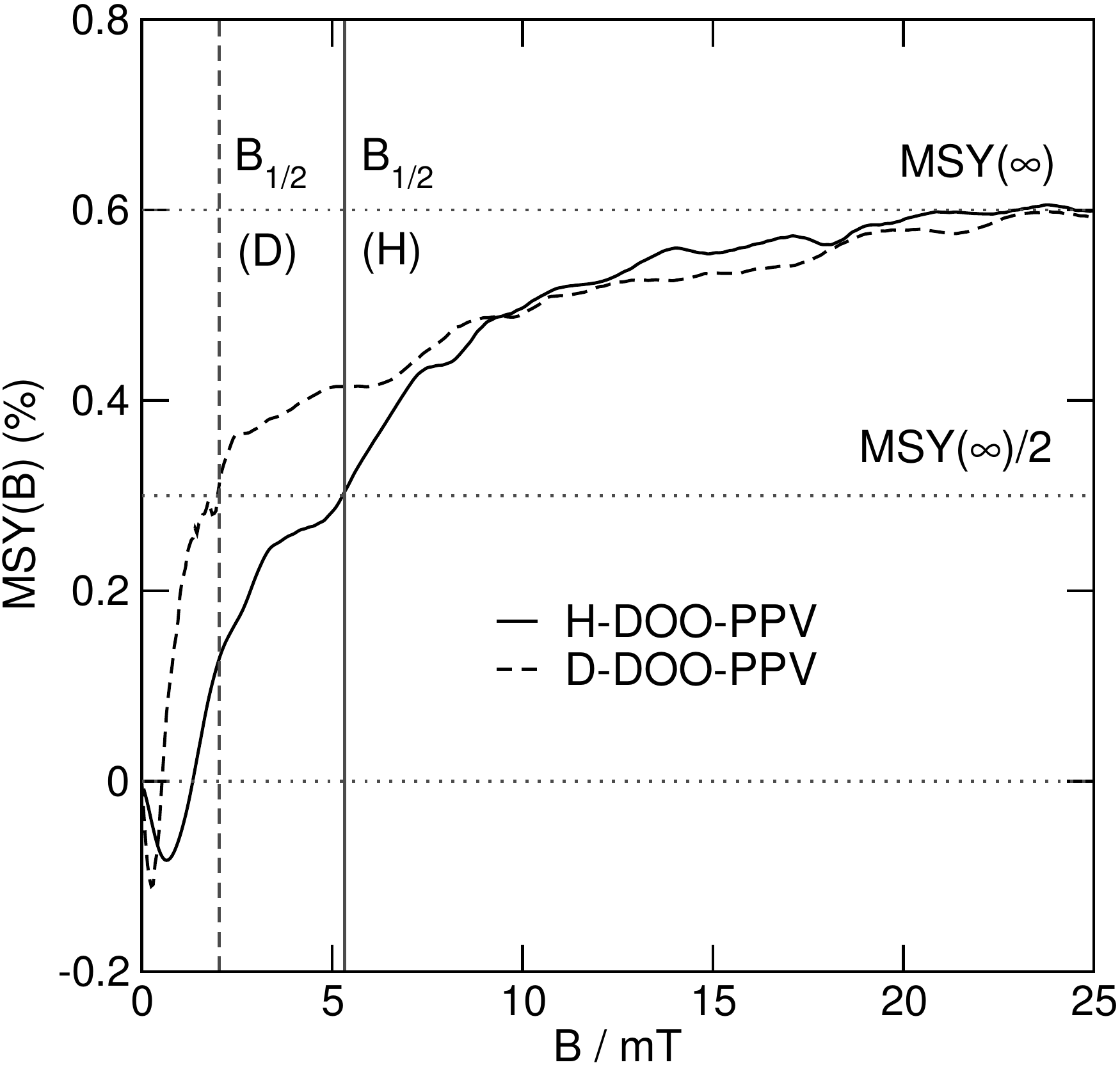}}
\caption{The magnetic field effects in the singlet yields of the polaron pair recombination reactions of H-DOO-PPV and D-DOO-PPV in Figs.~3 and 4, on an expanded scale that emphasises the difference between $B_{1/2}({\rm H})$ and $B_{1/2}({\rm D})$ (the magnetic field strengths at which the magnetic field effect on the singlet yield is half its asymptotic value MSY($\infty$)).}
\label{SYs}
\end{figure}

This suggests that another process plays an important role in the spin dynamics of the polaron pairs. Many other processes have already been considered and discounted,\cite{Ehrenfreund12} but comparatively little attention has yet been given to electron spin relaxation. This is perhaps unsurprising, because the most common mechanism of relaxation, the modulation of hyperfine interactions, leads to irrelevantly long spin-lattice relaxation times for the polarons in solid state organic polymers.\cite{Yang07}  However, other relaxation mechanisms could well be significant, such as the modulation of the exchange interaction between the two electrons in a polaron pair. The strength of this interaction depends exponentially on the separation between the two electrons, which varies due to the migration of the polarons within the pair.\cite{Willard15} It has been shown in other contexts that this modulation can cause the decay of coherences between the singlet and triplet states (``singlet-triplet dephasing"),\cite{Shushin91,Maeda12} and moreover that this can lead to $B_{1/2}$ values significantly larger than those predicted by the Weller equation.\cite{Maeda12}

In order to include this phenomenon in our semiclassical simulations, we can simply add an appropriate term to the right-hand side of Eq.~(32),\cite{Shushin91,Sowa15}
\begin{equation}
-\hat{\hat{L}}(B,{\bf h}_1,{\bf h}_2)\hat{\rho} = \cdots - k_{\rm R}\left(\hat{P}_{\rm S}\hat{\rho}\hat{P}_{\rm T}+\hat{P}_{\rm T}\hat{\rho}\hat{P}_{\rm S}\right),
\end{equation}
where $k_{\rm R}$ is a singlet-triplet dephasing rate constant and $\hat{P}_{\rm S}$ and $\hat{P}_{\rm T}$ are the singlet and triplet projection operators in Eqs.~(10) and~(11). This extra (manifestly hermitian) term clearly results in the decay of coherences between the singlet and triplet states while leaving their populations unchanged (as required).

It should be noted that the average exchange interaction has been neglected in the Hamiltonian in Eq.~(30), even though its modulation is included here. There are several reasons for this. Firstly, the average strength of the exchange coupling constant is extremely difficult to determine in these organic polymers,\cite{Lane97} and we are reluctant to introduce any more arbitrary parameters into our calculations than are absolutely necessary. Secondly, the exponential dependence of the exchange interaction on distance implies that even if the average exchange interaction were negligible, rare excursions to short distances might well be enough to give a significant modulation effect. And thirdly, we have found in exploratory calculations that the spin dynamics are largely unaffected by the inclusion of a physically reasonable average exchange coupling. We shall therefore neglect this average coupling for simplicity. 

\section{Simulations of DOO-PPV}

\begin{figure}[t]
\centering
\resizebox{0.9\columnwidth}{!} {\includegraphics{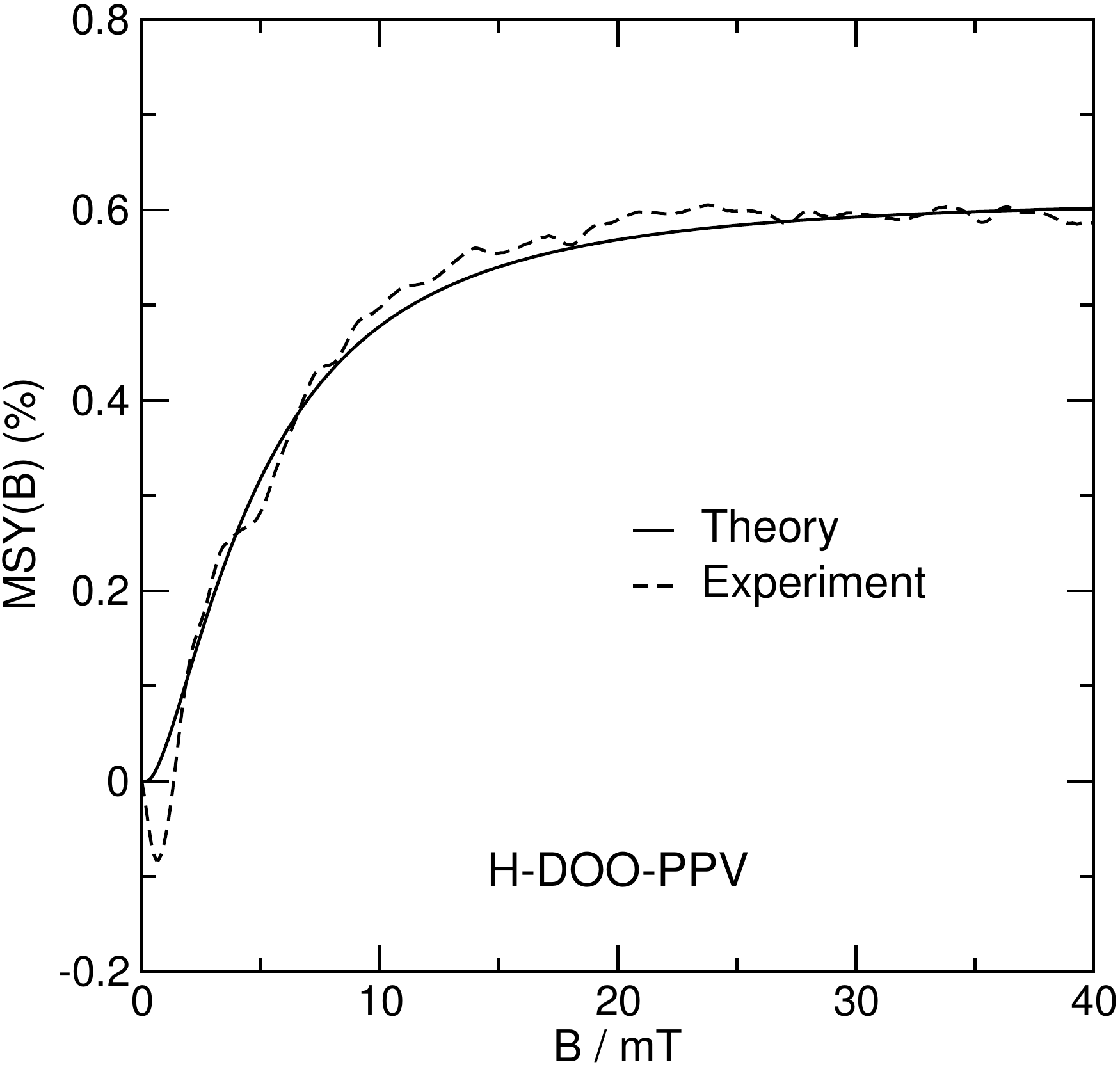}}
\caption{Experimentally determined and theoretically calculated magnetic field effects in the singlet yield of H-DOO-PPV. The experimental curve was obtained from Eq.~(24) using the magnetoelectroluminescence and magnetoconductance data reported in Refs.~\onlinecite{Nguyen10a} and~\onlinecite{Nguyen10b}.}
\label{H-SY}
\end{figure}

With singlet-triplet dephasing included, we are finally in a position to simulate the experimental results in Figs.~3 and~4 and see how well the simulations match the experiments. Since we have already specified what we believe to be realistic hyperfine fields $B_{\rm hyp}({\rm H})$ and $B_{\rm hyp}({\rm D})$ for the undeuterated and deuterated DOO-PPV polarons, the simulation only involves three free parameters: the overall singlet and triplet polaron pair decay rate constants $k_{\rm S}$ and $k_{\rm T}$ and the singlet-triplet dephasing rate constant $k_{\rm R}$.  We have therefore performed a combined least squares fit to the experimental data for both polymers in the $\{k_{\rm R},k_{\rm S},k_{\rm T}\}$ parameter space using the downhill simplex method, which was found to yield a single minimum at $k_{\rm R}=1.50\times 10^8$ s$^{-1}$, $k_{\rm S}=4.84\times 10^5$ s$^{-1}$, and $k_{\rm T}=4.94\times 10^5$ s$^{-1}$.

\begin{figure}[t]
\centering
\resizebox{0.9\columnwidth}{!} {\includegraphics{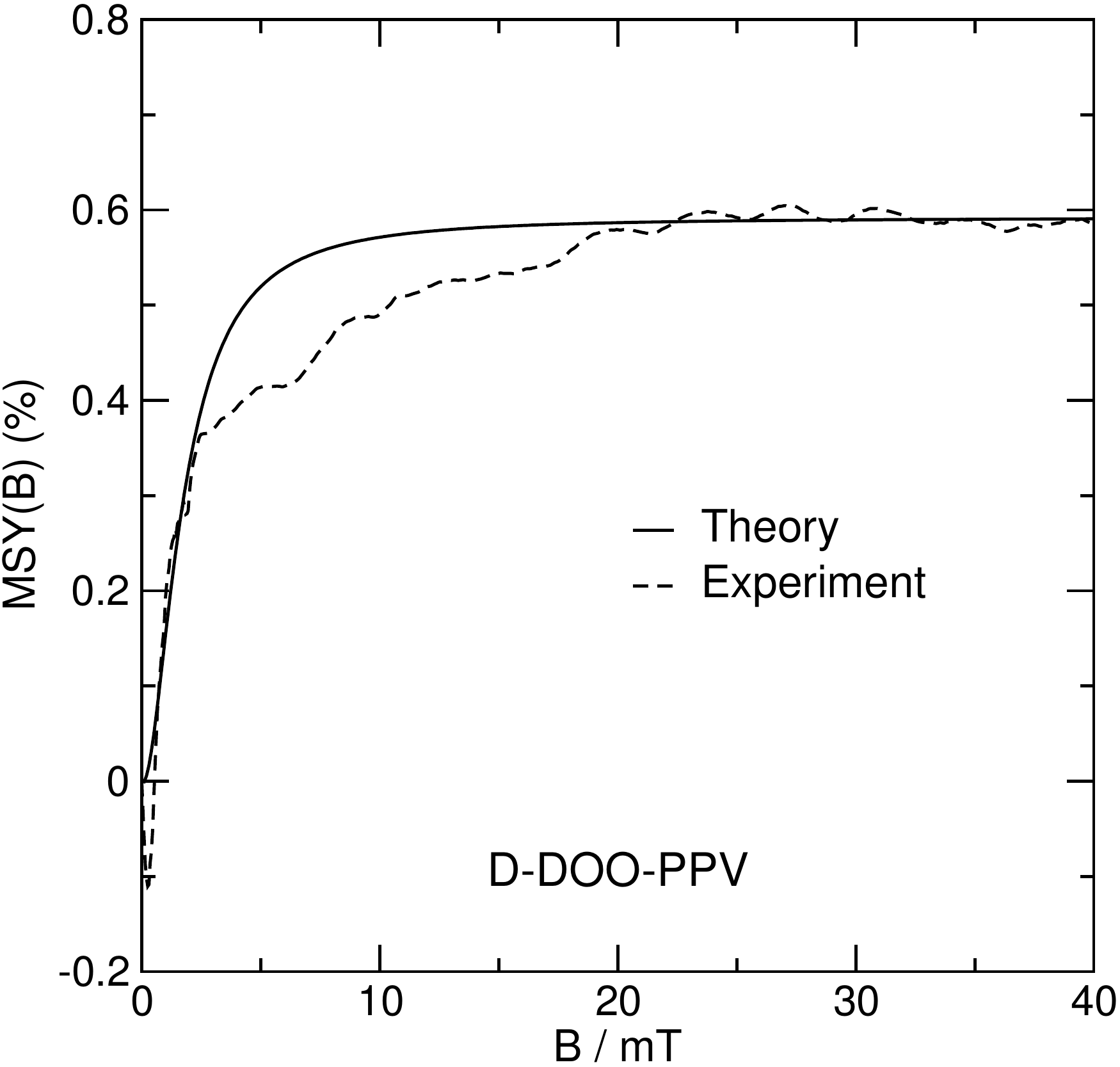}}
\caption{As in Fig.~6, but for D-DOO-PPV. The experimental data is again from Refs.~\onlinecite{Nguyen10a} and~\onlinecite{Nguyen10b}.}
\label{D-SY}
\end{figure}

The resulting theoretical curves are compared with the MSY($B$) curves obtained from the magnetoelectroluminescence and magnetoconductance measurements of Nguyen {\em et al.}\cite{Nguyen10a, Nguyen10b} in Figs.~6 and 7. The agreement between theory and experiment is clearly excellent for H-DOO-PPV (Fig.~6) at all but the very lowest magnetic field strengths. The agreement is also good for D-DOO-PPV (Fig.~7), although the experimental curve does have a kink between  $B=2.5$ and 20 mT that is not captured by the theoretical calculation. Assuming that this can be dismissed as an artefact, the agreement between theory and experiment is clearly very good for both isotopologues of DOO-PPV over a wide range of applied magnetic fields. 

This both justifies the expression for MEL($B$) in Eq.~(24) and shows that singlet-triplet dephasing plays an important role in the polaron pair spin dynamics. When the same three parameters $k_{\rm R}$, $k_{\rm S}$ and $k_{\rm T}$ are optimised to fit the computed MSY($B$) to the experimental MEL($B$) as has been done in the past\cite{Ehrenfreund12,Nguyen10a,Kersten11} (see Eq.~(5)), the resulting fits for the two isotopologues are nowhere near so compelling as those in Figs.~6 and~7.  And when singlet-triplet dephasing is switched off by setting $k_{\rm R}=0$, it is not possible to reproduce the $B_{1/2}$ values of the experimental MEL($B$) curves without invoking unfeasibly large and unrelated hyperfine fields $B_{\rm hyp}({\rm H})$ and $B_{\rm hyp}({\rm D})$ (see the discussion of the Weller equation in Sec.~VII).  

One feature of the experimental results that is not reproduced by our calculations is the low field effect (LFE),\cite{Timmel98} also referred to as the ultra-small magnetic field effect (USMFE),\cite{Ehrenfreund12,Nguyen13} which results in a negative dip in MSY($B$) below $1.3$ mT for H-DOO-PPV and $0.5$ mT for H-DOO-PPV. This LFE has been captured in earlier calculations,\cite{Ehrenfreund12,Nguyen10a,Kersten11} and we too have found that it can be captured by optimising the parameters $k_{\rm R}$, $k_{\rm S}$ and $k_{\rm T}$ to fit just the low field region. However, the resulting MSY($B$) curves do not give nearly such good agreement with experiment over the full range of magnetic field strengths as those in Figs.~6 and~7. 

It should be noted that the Schulten-Wolynes approximation is expected to be least reliable in the low field region. In fact, it has been shown to give a quantitatively incorrect LFE in model calculations on a moderately sized radical pair for which the exact quantum mechanical calculation could be performed for comparison.\cite{Manolopoulos13} We do not therefore feel that the lack of a (small) LFE in the present calculations is worth pursuing further here. This effect could be captured more accurately by replacing the Schulten-Wolynes approximation with an improved semiclassical theory in which the nuclear spin precession is taken into account.\cite{Manolopoulos13,Lewis14} However that would require a detailed knowledge of the hyperfine coupling constants in the two polarons which we do not have, and which the Schulten-Wolynes approximation conveniently avoids.

\section{Concluding Remarks}

While the mechanism of the magnetoconductance in organic polymer films is still somewhat controversial,\cite{Frankevich92,Bobbert07,Cox14,Lupton08,Hu07,Janssen11} the magnetoelectroluminescence is widely accepted to arise from the polaron pair recombination scheme illustrated in Fig.~1. In this paper, we have used this scheme to derive the expression in Eq.~(24) relating the theoretically calculable magnetic field effect on the singlet yield of the polaron pair recombination reaction, MSY$(B)$, to the experimentally measurable magnetoelectroluminescence MEL($B$) and magnetoconductance MC$(B)$ of the polymer film. We have argued that this expression holds independently of the mechanism of the magnetoconductance, provided the mobilities of the positive and negative polarons are not affected by the  magnetic field {(i.e., provided there is no contribution from spin-dependent transport).}

We have also discussed the semiclassical calculation of MSY$(B)$ for polymers such as DOO-PPV, the standard deviations of the hyperfine fields in which can be extracted from ESR linewidth measurements.\cite{Kuroda00,Zezin04}  Once these hyperfine fields are known, the semiclassical (Schulten-Wolynes\cite{Schulten78}) calculation of MSY($B$) involves just three empirical parameters: the overall singlet and triplet decay rate constants $k_{\rm S}$ and $k_{\rm T}$ of the polaron pair and a singlet-triplet dephasing rate constant $k_{\rm R}$. The last of these parameters has not been included in any previous theory of magnetoelectroluminescence that we are aware of, but we have argued on the basis of the Weller equation\cite{Weller83} that its inclusion is essential to reproduce the experimental results of Nguyen {\em et al.}\cite{Nguyen10a,Nguyen10b} for H-DOO-PPV and D-DOO-PPV.

The fits of the resulting theory to the experimental data are shown in Fig.~6 for H-DOO-PPV and Fig.~7 for D-DOO-PPV. The agreement between theory and experiment is clearly very good for both isotopologues over a wide range of magnetic field strengths. The empirical parameters obtained from the fit are $k_{\rm R}=1.50\times 10^8$ s$^{-1}$, $k_{\rm S}=4.84\times 10^5$ s$^{-1}$ and $k_{\rm T}=4.94\times 10^5$ s$^{-1}$, and all that remains is to explain why we feel that these parameters are reasonable. 

Dividing $k_{\rm R}$ by the gyromagnetic ratio of the electron, $\gamma_{\rm e}=1.76\times 10^{11}$ s$^{-1}$T$^{-1}$, we find that the singlet-triplet dephasing can be associated with a magnetic field of 0.85 mT, which is somewhat larger than the hyperfine field in either isotopologue ($B_{\rm hyp}({\rm H})\simeq 0.37$ mT and $B_{\rm hyp}({\rm D})\simeq 0.093$ mT). This is entirely consistent with our discussion of the Weller equation in Sec.~VII, and it suggests that singlet-triplet dephasing does indeed play a significant role in determining the experimentally-observed $B_{1/2}$ values for the two isotopologues. 

\begin{figure}[t]
\centering
\resizebox{0.9\columnwidth}{!} {\includegraphics{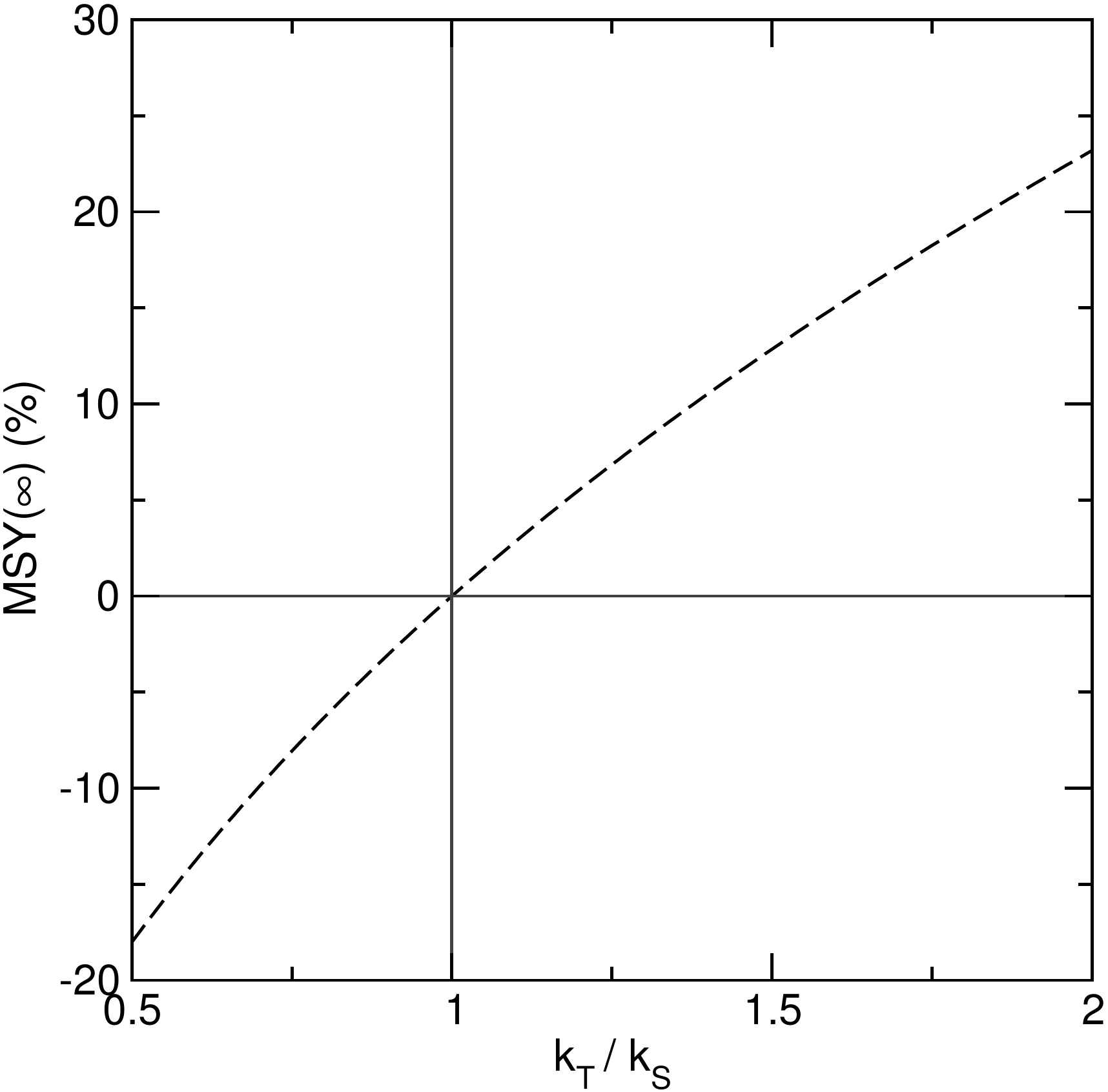}}
\caption{Plot of the computed MSY($\infty$) versus $k_{\rm T}/k_{\rm S}$ for H-DOO-PPV, with $k_{\rm R}$ and $k_{\rm S}$ as in Figs.~6 and~7.}
\label{D-SY}
\end{figure}

Our singlet and triplet polaron pair decay rate constants $k_{\rm S}$ and $k_{\rm T}$ are similar, and they have a reasonable magnitude, corresponding to a polaron pair lifetime of a couple of microseconds. The only aspect of these rates that needs further comment is the fact that $k_{\rm S}$ is smaller than $k_{\rm T}$. This is at odds with at least one theoretical treatment of singlet and triplet exciton formation rates from polaron pairs.\cite{Barford04} However, it is an inevitable consequence of the experimental data in Figs.~6 and~7: a positive magnetic field effect in the high-field limit, MSY$(\infty)>0$, can only be obtained when $k_{\rm T}>k_{\rm S}$.\cite{Kersten11} This is shown graphically in Fig.~8, which plots the computed MSY($\infty$) as a function of $k_{\rm T}/k_{\rm S}$ for H-DOO-PPV when $k_{\rm R}$ and $k_{\rm S}$ have the same values as in Figs.~6 and~7.

The behaviour in Fig.~8 can be explained as follows. When $k_{\rm T}=k_{\rm S}$, Eqs.~(15) and~(33) both give $\Phi_{\rm S}(B)=1/4$ for all $B$, and therefore MSY($\infty$) = 0. But when $k_{\rm T}>k_{\rm S}$, the triplet states of the polaron pair decay more rapidly than the singlet state, and $\Phi_{\rm S}(B)$ is less than 1/4. The decrease in $\Phi_{\rm S}(B)$ is most pronounced at low field strengths, where singlet population can be lost through intersystem crossing to all three components (T$_0$, T$_{+1}$, T$_{-1}$) of the triplet. At high field strengths the Zeeman splittings of the T$_{\pm 1}$ triplet components make them energetically inaccessible to singlet-triplet interconversion, so singlet population can only be lost through interconversion to T$_0$. Therefore $\Phi_{\rm S}(0)<\Phi_{\rm S}(\infty)<1/4$ and MSY$(\infty)>0$. Conversely, when $k_{\rm T}<k_{\rm S}$, a similar argument gives $\Phi_{\rm S}(0)>\Phi_{\rm S}(\infty)>1/4$ and MSY$(\infty)<0$.

Finally, while we have deliberately avoided making any assumptions about the mechanism of the magnetoconductance in DOO-PPV, our results do in fact shed some light on this. One of the most widely discussed mechanisms of magnetoconductance is the polaron pair mechanism, in which the effect of the magnetic field on the conductance is assumed to arise from its effect on the dissociation yield of the polaron pair back to free charge carriers (P$^+$ and P$^-$).\cite{Ehrenfreund12,Frankevich92} However, it is clear from Figs.~3 and~4 that the magnetic field effect on the overall singlet yield of the polaron pair, which includes both the singlet exciton yield and the yield of free charge carriers in the singlet state, has already saturated at a field strength of 20 mT, whereas the  magnetoelectroluminescence and magnetoconductance continue to increase at higher field strengths (up to 40 mT). Since when the overall singlet yield has saturated the overall triplet yield (including the yield of free charge carriers in the triplet state) must have saturated too, this clearly implies that the change in the magnetoconductance beyond 20 mT cannot come from the polaron pair mechanism. {So although it seems likely, given the similarities between the MC($B$) and MSY($B$) curves in Figs.~3 and~4, that the polaron pair mechanism of magnetoconductance does play an important role in DOO-PPV, there must also be some other mechanism in operation at high magnetic field strengths.}

\begin{acknowledgements}
We are grateful to Tho Nguyen for providing us with the experimental data from Refs.~\onlinecite{Nguyen10a} and \onlinecite{Nguyen10b}, and to him, William Barford, and Richard Friend for helpful discussions. This work was funded by the European Research Council under the European Union's 7th Framework Programme, FP7/2007-2013/ERC grant agreement no.~340451.
\end{acknowledgements}


\begin{thebibliography}{99}

\bibitem{Yang97}
Y.~Yang, Mater. Res. Soc. Bull. {\bf 22}, 31 (1997).

\bibitem{Hoofman98}
R.~J.~O.~M.~Hoofman, M.~P.~de Haas, L.~D.~A.~Siebbeles and J.~M.~Warman, Nature {\bf 392}, 54 (1998).

\bibitem{Grozema02}
F.~C.~Grozema, L.~P.~Candeias, M.~Swart, P.~Th.~van Duijnen, J.~Wildeman, G.~Hadziioanou, L.~D.~A.~Siebbeles, and J.~M.~Warman, J. Chem. Phys. {\bf 117}, 11366 (2002).

\bibitem{Lupton10}
J. M. Lupton, D. R. McCamey and C. Boehme, Chem. Phys. Chem., {\bf 11}, 3040 (2010).

\bibitem{Ehrenfreund12}
E.~Ehrenfreund, Z.~V.~Vardeny, Isr. J. Chem. {\bf 52}, 552 (2012).

\bibitem{Nguyen10a}
T.~D.~Nguyen, G.~Hukic-Markosian, F.~J.~Wang, L.~Wojcik, X.~G.~Li, E.~Ehrenfreund, Z.~V.~Vardeny, Nat. Mater. {\bf 9}, 345 (2010).

\bibitem{Nguyen10b}
T.~D.~Nguyen, B.~R.~Gautam, E.~Ehrenfreund, Z.~V.~Vardeny, Phys. Rev. Lett. {\bf 105}, 166804 (2010).

\bibitem{Kersten11}
S. P. Kersten, A. J. Schellekens, B. Koopmans and P. A. Bobbert, Phys. Rev. Lett. {\bf 106}, 197402 (2011).

\bibitem{Frankevich92}
E. L. Frankevich, A. A. Lymarev, I. Sokolik, F. E. Karasz, S. Blumstengel, R. H. Baighman and H. H. H\"orhold, Phys. Rev. B {\bf 46}, 9320 (1992).

\bibitem{Hu07}
B. Hu and Y. Wu, Nat. Mater. {\bf 6}, 985 (2007).

\bibitem{Bobbert07}
P. A. Bobbert, T. D. Nguyen, F. W. A. van Oost, B. Koopmans and M. Wohlgenannt, Phys. Rev. Lett. {\bf 99}, 216801 (2007).

\bibitem{Lupton08}
J. M. Lupton and C. Boehme, Nat. Mater. {\bf 7}, 598 (2008).

\bibitem{Janssen11}
P. Janssen, W. Wagemans, W. Verhoeven, E.H.M. van der Heijden, M. Kemerink and B. Koopmans, Synth. Metals, {\bf 161}, 617 (2011).

\bibitem{Cox14}
M. Cox, M. H. A. Wijnen, G. A. H. Wetzelaer, M. Kemerink, P. W. M. Blom and B. Koopmans, Phys. Rev. B {\bf 90}, 155205 (2014).

\bibitem{Schulten78}
K.~Schulten and P.~G.~Wolynes, J. Chem. Phys. {\bf 68}, 3292 (1978).

\bibitem{Kuroda00}
S. Kuroda, K. Marumoto, H. Ito, N. C. Greenham, R.H. Friend, Y. Shimoi and S. Abe, Chem. Phys. Lett. {\bf 325}, 183 (2000).

\bibitem{Zezin04}
A. A. Zezin, V. I. Feldman, J. M. Warman, J. Wildeman and G. Hadziioannou, Chem. Phys. Lett. {\bf 389}, 108 (2004).

\bibitem{Weller83}
A. Weller, F. Nolting and H. Staerk, Chem. Phys. Lett. {\bf 96}, 24 (1983).

\bibitem{Pope99}
M. Pope and C. E. Swenberg, {\em Electronic Processes in Organic Crystals and Polymers}, 2nd Ed., Oxford University Press, 1999.

\bibitem{Breuer06}
H-P. Breuer and F. Petruccione, {\em The Theory of Open Quantum Systems}, Clarendon Press, Oxford, 2006.

\bibitem{Willard15}
P. B. Deotare, W. Chang, E. Hontz, D. N. Congreve, L.Shi, P. D. Reusswig, B. Modtland, M. E. Bahlke, C. K. Lee, A. P. Willard, V. Bulovic, T. Van Voorhis and M. A. Baldo, Nat. Mater. {\bf 14}, 1130 (2015).

\bibitem{Haberkorn76}
R.~Haberkorn, Mol.~Phys. {\bf 32}, 1491 (1976).

\bibitem{Jones10}
J. A. Jones and P. J. Hore, Chem. Phys. Lett. {\bf 488}, 90 (2010).

\bibitem{Parmenter59}
R.~H.~Parmenter and W.~Ruppel, J. Appl. Phys. {\bf 30}, 1548 (1959).

\bibitem{Leger03}
J.~M.~Leger, S.~A.~Carter, B.~Ruhstaller, H.-G.~Nothofer, U.~Scherf, H.~Tilman and H.-H.~H\"orhold, Phys. Rev. B {\bf 68}, 054209 (2003).

\bibitem{MELapproximation}
The approximation in Eq.~(20) neglects relative terms of order $[\delta n(B)/n(B)]^2$.

\bibitem{MCapproximation}
The approximation in Eq.~(22) neglects relative terms of order $\delta n(B)/n(B)$ unless $\mu_+=\mu_-$, in which case it becomes exact.

\bibitem{Marcus14}
M. Marcus, O. R. Tozer and W. Barford, J. Chem. Phys. {\bf 141}, 164102 (2014).

\bibitem{Shimoi95}
Y.~Shimoi, S.~Abe, S.~Kuroda and K.~Murata, Solid State Commun. {\bf 95}, 137 (1995).

\bibitem{ParticleHole}
Particle-hole symmetry will be broken by the electronegative oxygen atoms that are $\alpha$ to the conjugated framework of DOO-PPV. However, explicit density functional theory calculations on short oligomers of a more manageable compound suggest that this is only a minor effect:  the computed hyperfine field strengths $B_{{\rm hyp},i}$ of the positively and negatively charged $n=3$ oligomers of DOE-PPV (where E = ethyl) differ by less than 10\%.

{\bibitem{Ratioof4}
The factor of 1/4 in Eq.~(35) would be obtained if D-DOO-PPV were fully deuterated. However, the material used in Refs.~6 and~7 was only partially deuterated: the H atoms directly bonded to the conjugated $\pi$ system were deuterated whereas those in the DOO side chains were not. The question that then arises is whether the nuclear spins in the side chains contribute to $B_{\rm hyp}$. In fact, the explicit DFT calculations on $n=3$ oligomers of DOE-PPV mentioned in Ref.~30 indicate that the hyperfine couplings in the side chains make up less than 0.25\% of $B_{\rm hyp}$, as one would expect on the basis of chemical intuition. Therefore, while strictly speaking the ratio $B_{\rm hyp}({\rm H})/B_{\rm hyp}({\rm D})$ will be slightly less than 4, it will certainly not be as small as the ratio $B_{1/2}({\rm H})/B_{\rm 1/2}({\rm D}) \simeq 2.65$ discussed in Sec.~VII.} 

\bibitem{Yang07}
C.~G.~Yang, E.~Ehrenfreund and Z.~V.~Vardeny, Phys. Rev. Lett. {\bf 99}, 157401 (2007).

\bibitem{Shushin91}
A. I. Shushin, Chem. Phys. Lett. {\bf 181}, 274 (1991).

\bibitem{Maeda12}
K.~Maeda, A.~J.~Robinson, K.~B.~Henbest, H.~J.~Hogben, T.~Biskup, M.~Ahmad, E.~Schleicher, S. Weber, C.~R.~Timmel and P.~J.~Hore, Proc. Natl. Acad. Sci. USA {\bf 109}, 4774 (2012).

\bibitem{Sowa15}
J.~K.~Sowa, {\em The Role of Dephasing in Radical Pair Magnetoreceptors}, Part II Chemistry Thesis, Oxford University, 2015.

\bibitem{Lane97}
P. A. Lane, X. Wei and Z. V. Vardeny, Phys. Rev. B, {\bf 56}, 4626 (1997).

\bibitem{Timmel98}
C.~R.~Timmel, U.~Till, B.~Brocklehurst, K.~A.~McLauchlan and P.~J.~Hore, Mol. Phys. {\bf 95}, 71 (1998).

\bibitem{Nguyen13}
T. D. Nguyen, E. Ehrenfreund and Z. V. Vardeny, Org. Elec. {\bf 14}, 1852 (2013).

\bibitem{Manolopoulos13}
D.~E.~Manolopoulos and P.~J.~Hore, J. Chem. Phys. {\bf 139}, 124106 (2013).

\bibitem{Lewis14}
A.~M.~Lewis, D.~E.~Manolopoulos and P.~J.~Hore, J. Chem. Phys. {\bf 141} 044111 (2014).

\bibitem{Barford04}
W. Barford, Phys. Rev. B {\bf 70}, 205204 (2004).

\end{thebibliography}
\end{document}